\documentclass[11pt]{article}
\usepackage{moriond,epsfig}

\def\Journal#1#2#3#4{{#1} {\bf #2}, #3 (#4)}

\def\PLB{{\em Phys. Lett.}  B}

\def\be{\begin{equation}}
\def\ee{\end{equation}}
\def\bea{\begin{eqnarray}}
\def\eea{\end{eqnarray}}

\begin{document}
\vspace*{4cm}
\title{NON-SUSY SEARCHES AT THE TEVATRON}

\author{ M. EADS \footnote{On behalf of the CDF and D0 collaborations.}}

\address{Department of Physics and Astronomy, 116 Brace Lab,
University of Nebraska - Lincoln, USA }

\maketitle\abstracts{The CDF and D0 collaborations have results on a
large number of searches for beyond-the-standard-model phenomena. This
talk focuses on searches for non-supersymmetric model signatures. These
results, based on between 1--2.5 $fb^{-1}$ of data from $p \bar{p}$ 
collisions at the Fermilab Tevatron,
include some of the world's best limits on extra dimensions
models and heavy resonances.}

\section{Extra Dimension Searches}

Extra dimensions models have been proposed as a possible solution to the 
hierarchy problem. In ADD extra dimensions models (proposed by 
Arkani-Hamed, Dimopolous, and Dvali), standard model particles are
confined to the normal three spacial dimensions, while gravitons can
propagate in some number of extra dimensions \cite{add}.

In $p \bar{p}$ collisions, it is possible for a graviton to be produced 
in association with a photon. This graviton can then excape into the
extra dimensions, resulting in a detector signature of a single 
photon and missing transverse energy (MET). The D0 collaboration has results
on the search for this mono-photon signature based on 1~$fb^{-1}$ of 
data \cite{d0_monophoton}. This search requires a single photon in the
detector with transverse momentum ($p_T$) above 90~$GeV/c$ and MET above
70 $GeV$. The main backgrounds for this search are vector boson production
in association with a photon (where the $Z$ boson decays to neutrinos or
the charged lepton from the $W$ boson decay is lost), and instrumental 
backgrounds. Instrumental backgrounds include jets misidentifed as
photons, $W$ bosons decaying to electrons (when the electron is 
misidentified as a photon), cosmic rays, and beam halo events. After all
selections, a total of 29 events are observed in data, while 
$22.37 \pm 2.50$ background events are predicted. 95\% confidence
level (CL) limits on the fundamental Planck scale ($M_D$) are computed as 
a function of the number of extra dimensions. These limits vary 
from 884 $GeV/c^2$ for two extra dimensions to 779 $GeV/c^2$ for eight extra dimensions.

The CDF collaboration has also performed a search for ADD extra dimensions
by looking for events with a single photon and MET \cite{cdf_photonmet}.
This search uses 2 $fb^{-1}$ of data and requires a single photon with 
transverse energy ($E_T$) above 50 $GeV$ and MET greater than 90 $GeV$.
As in the D0 search, the dominant background is the production of a 
$Z$ boson in association with a photon, with the $Z$ boson decaying to
neutrinos. After all selections, 40 events are observed in data with
an expected bacground of $46.7 \pm 3.0$ events. 95\% CL limits on the 
fundamental Planck mass ($M_D$) are calculated that vary from 1080 $GeV/c^2$ for
two extra dimensions to 900 $GeV/c^2$ for six extra dimensions.

\begin{figure}
  \centering
  \psfig{figure=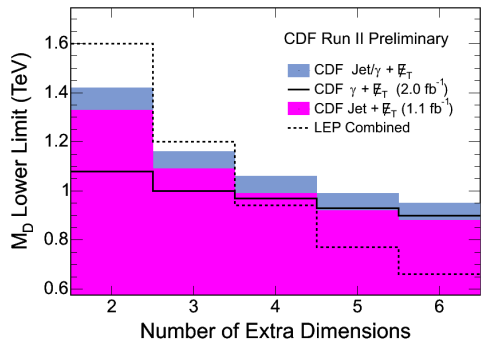,height=2in}
\caption{The 95\% CL lower limit on the fundamental Planck scale ($M_D$)
versus the number of extra dimensions for the CDF photon plus MET search,
jets plus MET search, and the combination.
\label{fig:cdf_ADD_limits}}
\end{figure}

ADD extra dimensions can also
produce a quark in association with a graviton. If the graviton excapes
into the extra dimensions, the detector signature will be jets in association
with MET. 
The CDF collaboration has also combined the photon plus MET results with
an analysis using jets with MET. 
The 95\% CL limits from the photon plus MET search 
combined with the jets plus MET search on $M_D$ vary from 1420 $GeV/c^2$ for
two extra dimensions to 950 $GeV/c^2$ for six extra dimensions. The limits
from the photon plus MET search, the jets plus MET search, and the combination 
are shown in Fig.\ \ref{fig:cdf_ADD_limits}.

Another model for extra dimensions is that of Randall and Sundrum
\cite{rs}. In Randall-Sundrum (RS) extra dimensions, standard model
particles are confined to a 3-brane and gravity originates on a 
different 3-brane. Only gravitons can propagate in the bulk between
these 3-branes. This model predicts that gravitons will appear as a
tower of Kaluza-Klein states.

The D0 collaboration has performed a search for RS gravitons decaying to
electrons or photons with 1 $fb^{-1}$ of data \cite{d0_rsgraviton_ee}. Events
are selected that contain two clusters of energy in the electromagnetic
calorimeter that have $p_T$ above 25 $GeV/c$. Since no track is required
in the central tracking system, this selection is sensitve to both electrons
and photons. The primary background is instrumental --- jets which are
misidentified as electromagnetic objects. Events are counted in a sliding
window in the invariant mass distribution of the two electronmagnetic objects.
The size of this window has been optimized from a simulation of the
detector response to RS gravitions with a range of masses. The observed number 
of events is consistent with the expected background, so 95\% CL limits 
are computed for the mass of the first Kaluza-Klein excitation of the graviton
($M_1$) as a function of coupling of the graviton to standard model
particles ($k/\overline{M}_{Pl}$). For $k/\overline{M}_{Pl}$ of 0.1, gravitons with
a mass below 900 $GeV/c^2$ are excluded. The CDF collaboration has 
performed a similar search in the dielectron channel 
with 2.5 $fb^{-1}$ of data that excludes 
gravitons with a mass below 850 $GeV/c^2$ for the same value of
$k/\overline{M}_{Pl}$ \cite{cdf_dielectron}.

RS gravitons can also decay to $Z$ bosons. The CDF collaboration has
performed a search for RS gravitons decaying to two $Z$ bosons, which in
turn decay to electrons, using 1.1 $fb^{-1}$ of data \cite{cdf_rsgraviton_zz}. 
Events are selected 
that contain four electrons. The large number of electrons in the final 
state makes the efficiency for identifying electrons critical for this 
search. Specialized electron reconstructed algorithms were implemented
that show a factor of 2--4 improvement over the standard algorithms. The
dominant background for this search is hadrons faking electrons. This
background was estimated using data events in the low-mass region, which 
is devoid of the expected signal. There are no events present after all
selections in the signal region, with $0.028 \pm 0.009 (stat.) \pm
0.011 (syst.)$ expected background events. 95\% CL limits on the production
cross section times branching ratio of gravitons to $Z$ bosons are computed
that vary from 4 -- 6 $pb$ for gravitons with a mass between 500 and
1000 $GeV/c^2$ and $k/\overline{M}_{Pl}$ of 0.1.

\section{Resonance Searches}

There are a large number of new physics models that predict massive
particles that would decay to quarks, leptons, or gauge bosons. Examples
of these models include Kaluza-Klein excitations in extra dimensions 
models (such as the RS gravitons discussed in the previous section), 
excited quarks or leptons, or new gauge bosons. Since there are many
possible models that predict a similar detector signature, these searches are
usually designed to be signature-based. That is, pairs of standard model
particles are selected (such as dileptons and dijets), and the invariant
mass distribution of the pair of standard model objects is examined for
deviations from the standard model prediction.

The CDF collaboration has performed a search for a high mass resonance in
dijet events with 1.13 $fb^{-1}$ of data \cite{cdf_dijet}. 
This search selects events with
jets with the the absolute value of rapidity less than 1.0 and a dijet mass 
above 180 $GeV/c^2$. At large invariant masses, the spectrum should be
smoothly falling. Any ``bump'' in this spectrum would be an indication of
new physics. The invariant mass distribution is fit with a smooth function.
The functional form was determined by Pythia and Herwig simulations, but
the fit parameters are determined from data. No deviation from the standard
model is observed, so limits are set on a variety of models. Excited quarks
with masses between 260 and 870 $GeV/c^2$ are excluded at 95\% CL. 
Limits are also set on the mass of new gauge bosons, referred
to as $W'$ and $Z'$ bosons, which are predicted in some new physics models
that contain additional gauge fields.
$W'$ bosons with masses between 280 and 840 $GeV/c^2$ and $Z'$ bosons
with masses between 320 and 740 $GeV/c^2$ are excluded. 
\footnote{This search assumed that the $W'$ and $Z'$ bosons would
have couplings similar to the standard model $W$ and $Z$ boson.}

\begin{figure}
  \centering
  \psfig{figure=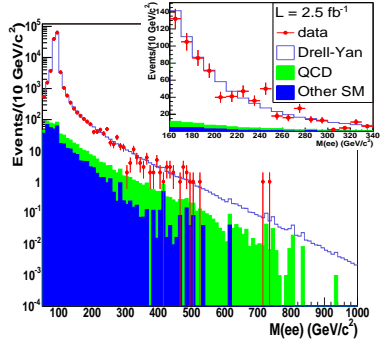,height=2.0in}
\caption{Invariant mass distribution for the CDF dielectron resonance search.
The inset shows the region around 240 $GeV/c^2$ in linear scale. 
\label{fig:cdf_dielectron_mass}}
\end{figure}

\begin{figure}
\begin{minipage}{0.5\textwidth}
  \centering
  \psfig{figure=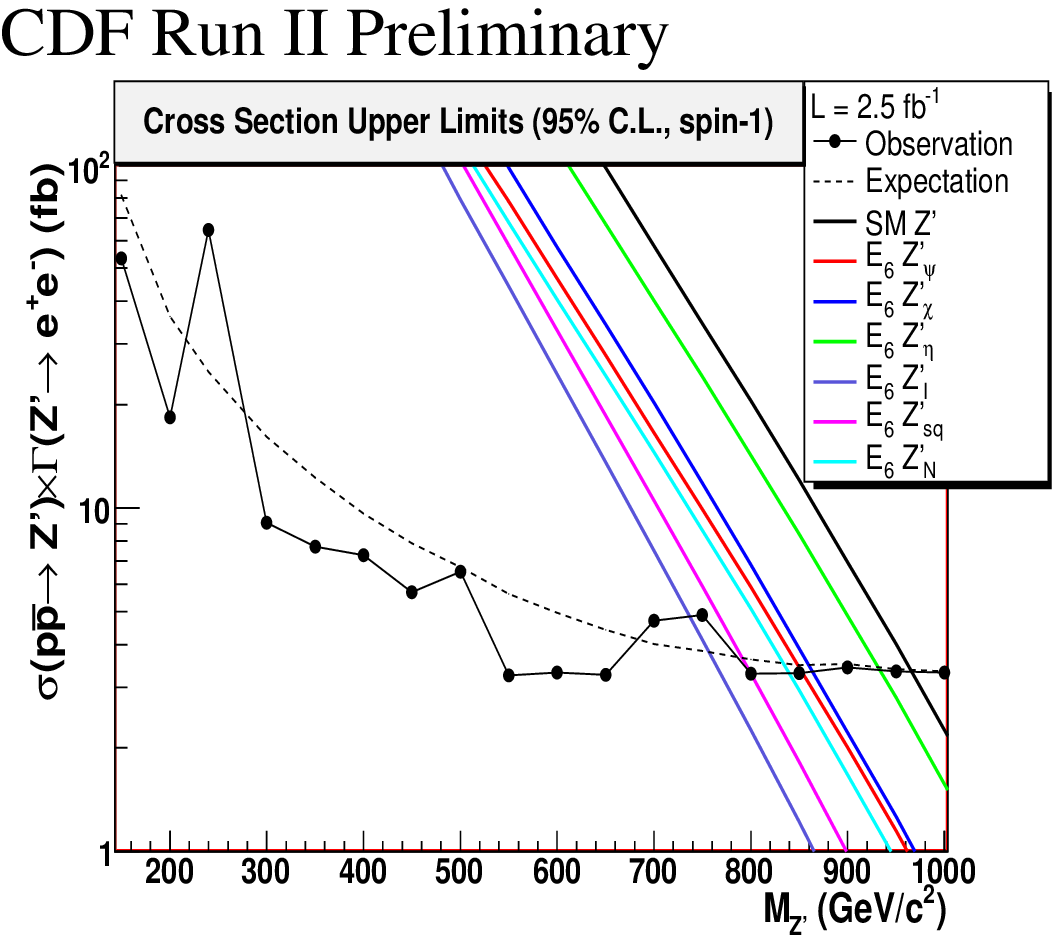,height=1.6in}
\end{minipage}
\begin{minipage}{0.5\textwidth}
  \centering
  \psfig{figure=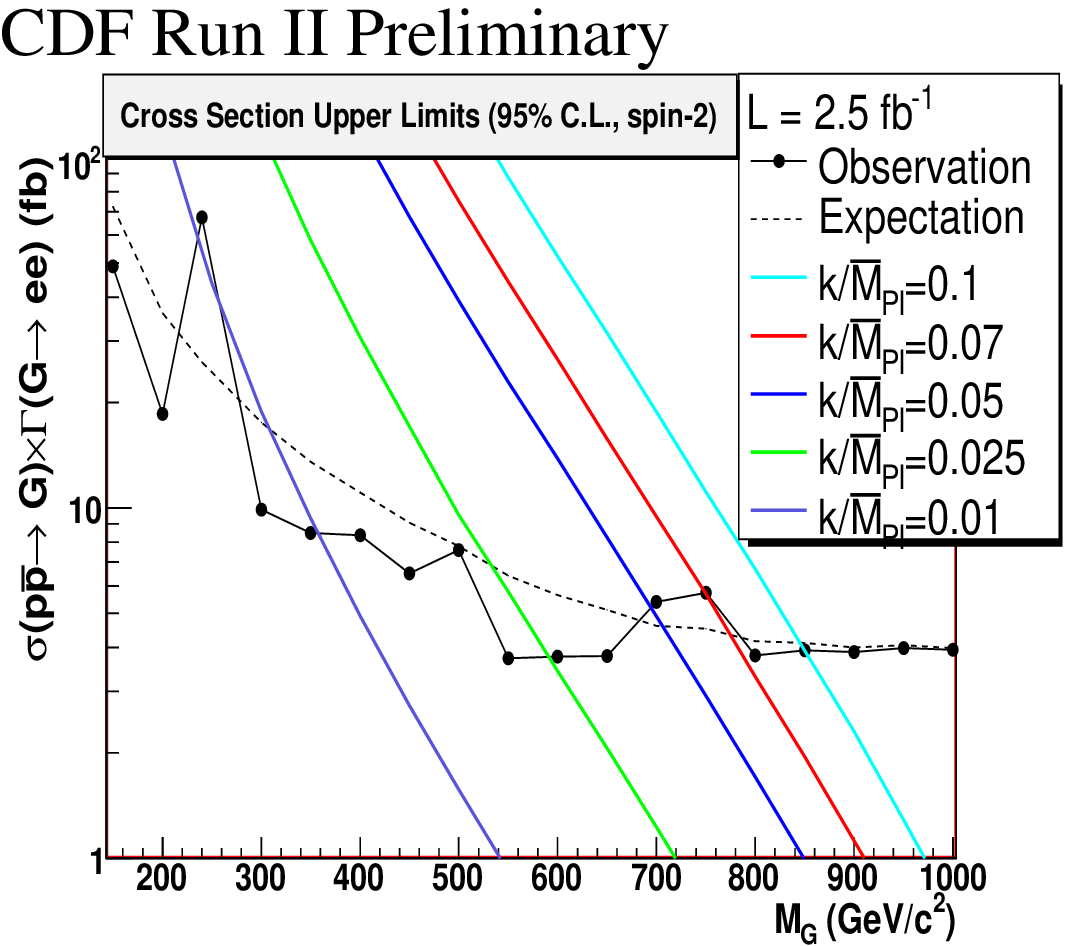,height=1.6in}
\end{minipage}
\caption{95\% CL limit from the CDF dielectron resosnance search for 
various $Z'$ bosons (left) and RS gravitons (right).
\label{fig:cdf_dielectron_limits}}
\end{figure}

The CDF collaboration has also performed a search for massive resonances
that decay to pairs of electrons, using 2.5 $fb^{-1}$ of data
\cite{cdf_dielectron}. This search requires two electrons with $E_T$ larger
than 25 GeV. The search region in the mass distriution is chosen to include invariant masses between
150 and 1000 $GeV/c^2$. The dominant, and irreducible, background is 
standard model Drell-Yan electron production. The shape of this background
is taken from detector simulations and the background is normalized to data in the
region of the $Z$ boson mass. There are also smaller backgrounds from 
standard model diboson production and instrumental backgrounds from
jets misidentified as electrons. The dielectron mass distribution is shown in 
Fig.\ \ref{fig:cdf_dielectron_mass}. The largest discrepency in an unbinned 
maximum likelihood fit for data events with invariant mass above 150 $GeV/c^2$ 
is a 3.8$\sigma$ excess in the range from 228 to 250 $GeV/c^2$.
Ensemble tests show that the probability of observing a discrepency of this 
size anywhere in the mass range from 150 to 1000 $GeV/c^2$ is 0.6\%. No
claim of new physics is made, and limits are set on a variety of models.
A $Z'$ boson (with standard model couplings) with a mass below 966 $GeV/c^2$
is excluded at 95\% CL. Limits are also set on RS gravitons (which were
previously discussed) and new $Z$ bosons in an $E_6$ model. The limits
on RS gravitons and $Z'$ bosons are shown in Fig.\ \ref{fig:cdf_dielectron_limits}.

\section{Conclusion} 

The CDF and D0 collaborations have performed a large number of searches 
for new physics in a wide variety of models. The results presented here
use up to 2.5 $fb^{-1}$ of data and represent some of the world's best
limits on extra dimensions models and heavy resonances. Although the results
presented here concentrated on extra dimensions models and heavy resonances,
both experiments have a large number of results on other new physics
models. Furthermore, both experiments
continue to collect data with high efficiency and so both improvements
to existing results and new searches can be expected in the future.

\end{document}